\documentstyle[aps,prl,amssymb,twocolumn,latexsym]{revtex}

\newcommand	{\beq}		 {\begin {equation}}
\newcommand	{\eeq}		 {\end	 {equation}}
\newcommand	{\ds} 	         {\displaystyle}
\newcommand	{\klm}[1]	  {{\, \left(\, {#1} \,\right) }}
\newcommand	{\ekl}[1]     {{\, \left[\, {#1} \,\right] }}
\newcommand	{\gkl}[1]     {{\, \left\{\, {#1} \,\right\}}}
\newcommand	{\skl}[1]     {{\left\langle {#1} \right\rangle}}

\newcommand{\ld}{\mbox{$l_{\mbox{\scriptsize D}}$}}
\newcommand{\lt}{\mbox{$l$}}
\newcommand{\lc}{\mbox{$l_c$}}
\newcommand{\lr}{\mbox{$l_r$}}
\newcommand{\li}{\mbox{$l_\ell$}}
\newcommand{\tw}{\widetilde{w}}

\begin{document}

\twocolumn[\hsize\textwidth\columnwidth\hsize\csname@twocolumnfalse\endcsname

\title{\bf Particle currents and the distribution of terrace sizes in unstable epitaxial 
growth}
\author{{\sc M. Biehl$^{1,2}$, M. Ahr$^{1}$, M. Kinne$^{3}$, W. 
Kinzel$^{1}$, 
   and S. Schinzer$^{1,2}$}} 
 \noindent
\address{
            $^{1}$Institut f\"ur Theoretische Physik und Astrophysik,
            $^{2}$Sonderforschungsbereich 410\\ 
            Julius--Maximilians--Universit\"at W\"urzburg\\
            Am Hubland, 97074 W\"urzburg, Germany\\
            $^{3}$Lehrstuhl f\"ur Physikalische Chemie II\\
            Friedrich--Alexander--Universit\"at Erlangen--N\"urnberg \\
            Egerlandstr. 3, 91058 Erlangen, Germany 
       }

\maketitle

\begin{abstract}
A solid--on--solid model of epitaxial growth in $1+1$ dimensions
is investigated in which slope dependent upward and downward
particle currents compete on the surface. The
microscopic mechanisms which give rise to these currents 
are the smoothening
incorporation of particles upon deposition and
an Ehrlich--Schwoebel barrier which hinders inter--layer transport 
at step edges.  We calculate the distribution of 
terrace sizes and the resulting currents on a stepped surface with  
a given inclination angle.  The cancellation of the competing effects
leads to the selection of a stable {\sl magic slope\/}. 
Simulation results are in very good agreement with the theoretical findings. 
\end{abstract}
{PACS numbers: 81.10.Aj, 05.70.Ln, 68.55.-a}  \ \\   ]

Epitaxial growth has become a standard method for the production
of high--quality crystals and films, needed for  e.g.\ semiconductor devices.
An overview of experimental techniques  can be found in \cite{Krim}, for instance.
Significant effort has been devoted to a theoretical 
understanding of the many morphologies and scaling behaviors 
that can be observed in epitaxial growth, see e.g.\ \cite{pimpinelli} for a review of 
theoretical approaches. 
Here we address the frequently observed phenomenon
of mounds in unstable growth  which has
attracted considerable interest, see e.g. 
\cite{rostkrug,siegert,tang,epl,sed,gitter}.
Specifically, we consider situations in which competing
smoothening and steepening effects control the surface morphology
and lead to the selection of a stable  slope
in the system.

We discuss potential microscopic mechanisms which result in the
emergence of mounds and slope selection in the frame of
a discrete $(1+1)$--dimensional model. 
The net particle currents as well as the distribution of terrace 
sizes on a surface of a given slope can be worked out and this 
allows then to evaluate the magic slope  as well as
the complete statistical properties of the emerging surface.
The analysis complements previous theoretical investigations 
which address the mean terrace size only or neglect 
fluctuations explicitly \cite{georg} in the spirit of Burton Cabrera Frank (BCF) 
theory \cite{pimpinelli,BCF}.
We demonstrate that the full distribution of terrace sizes
carries relevant information that should be taken into account. 
Our results suggest, for example,  that it should be possible
to identify relevant microscopic mechanisms  from 
experimental data. 

Our $(1+1)$--dimensional model obeys the solid--on--solid  (SOS)
restriction, 
i.e.\ the surface can be described by an integer array of height 
variables $h_k$. Single particles are  deposited at 
randomly chosen  sites $k \in \{1,2,\ldots,L\}$.
Upon arrival, an incorporation process moves a particle to
the lowest available site  within a neighborhood of $\pm 
R$  lattice  constants, i.e.\ the site $j$ with 
$h_j = \min \left\{h_{k-R},\ldots,h_k,\ldots
 h_{k+R} \right\} $. 
In case of a tie, the site closest to the deposition is  
chosen. Only if this is still ambiguous, an additional random 
selection is performed. Such a smoothening mechanism, in absence 
of further effects, is commonly associated with the Edwards--Wilkinson 
universality class of
growth \cite{pimpinelli}. 
The parameter $R={\cal O}(1)$ (in lattice constants) 
is termed the incorporation radius and sets the 
typical length
scale of the process. Various interpretations of incorporation have
been considered,  including downward funneling 
on non--trivial
lattices and knock--out--processes due to the          
momentum of incoming particles, see e.g.\  \cite{evans,yue}.

A particle which is, after deposition and possibly incorporation, 
not yet bound to a lateral neighbor diffuses on the surface by 
performing a random walk (RW) until it 
reaches an additional binding partner and
becomes immobile or until it collides with
another moving adatom and forms an island
nucleus. 
In a density of diffusing particles, this nucleation process
 would result in a typical collision free path 
$\ld$. In the case of irreversible aggregation on a flat
substrate, and if islands
of two or more adatoms are considered immobile, it has been
shown that  $\ld \propto \left(d/f\right)^{1/4}$ in $(1+1)$ 
dimensions \cite{pimpinelli,villain}. Here,  $d$ is the diffusion constant
and $f$ the incoming flux, with all 
lengths in dimensionless lattice constants.

In  \cite{epl,sed,gitter}  nucleation is represented in an 
effective single particle picture and $\ld$ fixes 
the typical distance of island nuclei in the first layers 
on a flat
substrate. 
After the formation of mounds, terrace sizes 
are much smaller than $\ld$, typically.
\iffalse
Throughout the following we will assume that typical terrace sizes
$\lt$ on the stepped surface are much smaller than the diffusion 
length
$\ld$.
\fi 
Hence we will completely neglect nucleation on the stepped surface
which is justified
for small incoming flux or fast diffusion, respectively.

Here, the RW ends whenever the particle sticks irreversibly to a lateral 
neighbor, i.e. when it reaches a terrace step. Attachment can occur 
from below and above, in principle, but this symmetry is broken due
to the so--called Ehrlich--Schwoebel (ES) effect \cite{pimpinelli}: 
An additional energy barrier $E_{es}$ at step edges hinders downward 
moves of diffusing adatoms as this would involve loosely bound 
intermediate positions.  Our model takes the ES effect into account by
assigning  a probability $ p_{es} \, \propto \exp\ekl{-E_{es}/(k T)}$ 
to
downward moves. 

The ES effect results in an  uphill current of adatoms
because particles will stick
to upper terraces preferably for any $p_{es} < 1$. 
On the other hand, incorporation 
constitutes a downhill
current. Both effects are slope dependent and their cancellation 
gives rise to
the formation of mounds with a well--defined inclination angle.
Once the structures have built up the {\it magic slope}, a coarsening 
process begins
which decreases the number of mounds, see \cite{epl,sed,gitter} for details.  
  
We will first work out the distribution of terrace sizes  
which emerges in our model on surfaces with a given, fixed 
inclination.
Further, we  will calculate 
the mean displacement of a particle's final position from
its deposition site. The latter corresponds to the net particle  
current on the growing surface and its zero as a function of the 
inclination angle determines the stable slope.

We consider a triple of terraces on an inclined surface
with a central terrace $c$ of width $l_{c}$ (lattice sites 
$j=1,2,\ldots l_c$)
and its neighbors $\ell$ to the left ($r$ to the right) with size $\li$ $(l_r)$.
Without loss of generality, we assume
that the surface height decreases to the right. 
For a particle deposited on a site $k$ of terrace $c$ with $l_c > R$ 
we have to distinguish the following cases:  \\[-6mm]
\begin{itemize} 
\item[(a)] $k > l_c - R$:~ the incorporation process places the 
particle  
    at its final position at site $l_c +1$, attaching to the lower 
terrace end.\\[-6mm]
\item[(b)] $1\leq k \leq l_c -R$:~ the particle performs a random 
walk until
 it reaches one of the {\it trap sites}  $j=1$ or $j=l_c+1$ where it 
comes to rest.  This includes deposition at site $k=1$ without 
subsequent 
diffusion. \\[-6mm]
\end{itemize}
A diffusing particle located at a site $j$  with $2\!\leq\!j\!\leq\!l_c\!-\!1$ moves
to one of the neighboring positions $j\pm1$ with equal probability 
$1/2$. 
The asymmetry of the RW (b) is due to the ES--barrier 
present for jumps from site $j=l_c$.  We denote 
with 
$p_{es}/2$ the
probability for a downward move to site $l_c+1$.
With probability 
$(1-p_{es})/2$ the move is
rejected and the particle remains at $j=l_c$ for the next time step, whereas with 
probability $1/2$ it jumps to $l_c -1$. 

A straightforward exercise
yields the probability
for  a RW  initiated at site $k$ to end in $j=\lc\!+\!1$ by downward 
diffusion:
$q(k,l_c) = \left. \klm{(k\!-\!1)\, p_{es}} \right/ \klm{1\!+\!(l_c 
\!-\!1)  p_{es}} $
which obviously satisfies $q(1,l_c)=0$ for all $p_{es}$.
For similar problems of this type see for instance
\cite{Shehawey}. 

Hence, the total probability $d(l_c)$ for a deposition event 
to occur on terrace $c$  with 
subsequent downward diffusion is $ d(\lc) = \Delta(\lc)/L$ 
 with   \\[-4mm]
\beq 
 \ds \label{deltadef}
 \Delta(\lc) = \sum_{k=1}^{l_c\!-\!R} q(k,\lc) \!=\! 
 \frac{\klm{\lc\!-\!R}\klm{\lc\!-\!R-1} p_{es}}
 {2 + 2 \klm{\lc-1} p_{es}} \eeq
 if $\lc > R$ and $\Delta(\lc) =0 $
  else.
 The second case accounts for the fact that any particle directly 
 deposited  onto a  terrace of width $\lc \leq R$ will be 
incorporated 
  without performing  diffusion.

 The quantity $\Delta(l_c)$ can be interpreted as the effective number of
 deposition sites which contribute to downward diffusion from terrace $c$. 
 The prefactor $1/L$ of $d(\lc)$ is simply the constant probability for deposition
 on any of the sites in the system.

 Now we can work out the probability $\tw(\lc\!\to\!\lc\!-\!1)$ for the  
 central terrace $c$ to be shortened by the next deposition event. 
 For $\lc >  R$ one finds 
  \begin{eqnarray}
   \tw (\lc\!\to\!\lc\!-\!1) &=& 
 \klm{R\!+\! [\lc\!-\!R\!-\!\Delta(\lc)] \!\
  +\!\Delta(\li)} / L  \nonumber \\
   &=&  \klm{\lc \!-\! \Delta(\lc) \!+\! \Delta(\li)} /L.
\label{shorten}
  \end{eqnarray}
  The first contribution, $R$, represents deposition on any of the $R$ sites
  left of terrace $c$. Note that the outcome of the subsequent 
  incorporation process is completely independent of the surface 
  configuration, in particular of the left neighbor terrace width 
  $\li$.  The second term $\ekl{\ldots}$ accounts for deposition on 
  $c$ with final attachment to the upper terrace. Finally 
  $\Delta(\li)/L$ is the probability for the shortening of $c$ through 
  diffusion from terrace $\ell$. 
  
 If $0<\lc<R$, only two processes can shorten the central terrace:
 incorporation from exactly $\lc$ sites left of $c$ and downward 
 diffusion from terrace $\ell$. One obtains $\tw (\lc\!\to\!\lc\!-\!1) =
 (\lc\!+\!\Delta(\li))/L$ and we finally observe
 that Eq.\ (\ref{shorten})
 is valid for any $\lc \geq 1$ since $\Delta(\lc)=0$ for all $\lc 
 \leq R$. Obviously, $\tw (\lc \to \lc-1) = 0$ if $\lc=0$ already.

\begin{figure}[t]
\begin{center}
\setlength{\unitlength}{2pt}
\begin{picture}(160,130)(0,0)
\put(0,0){\makebox(160,125)
          {\includegraphics{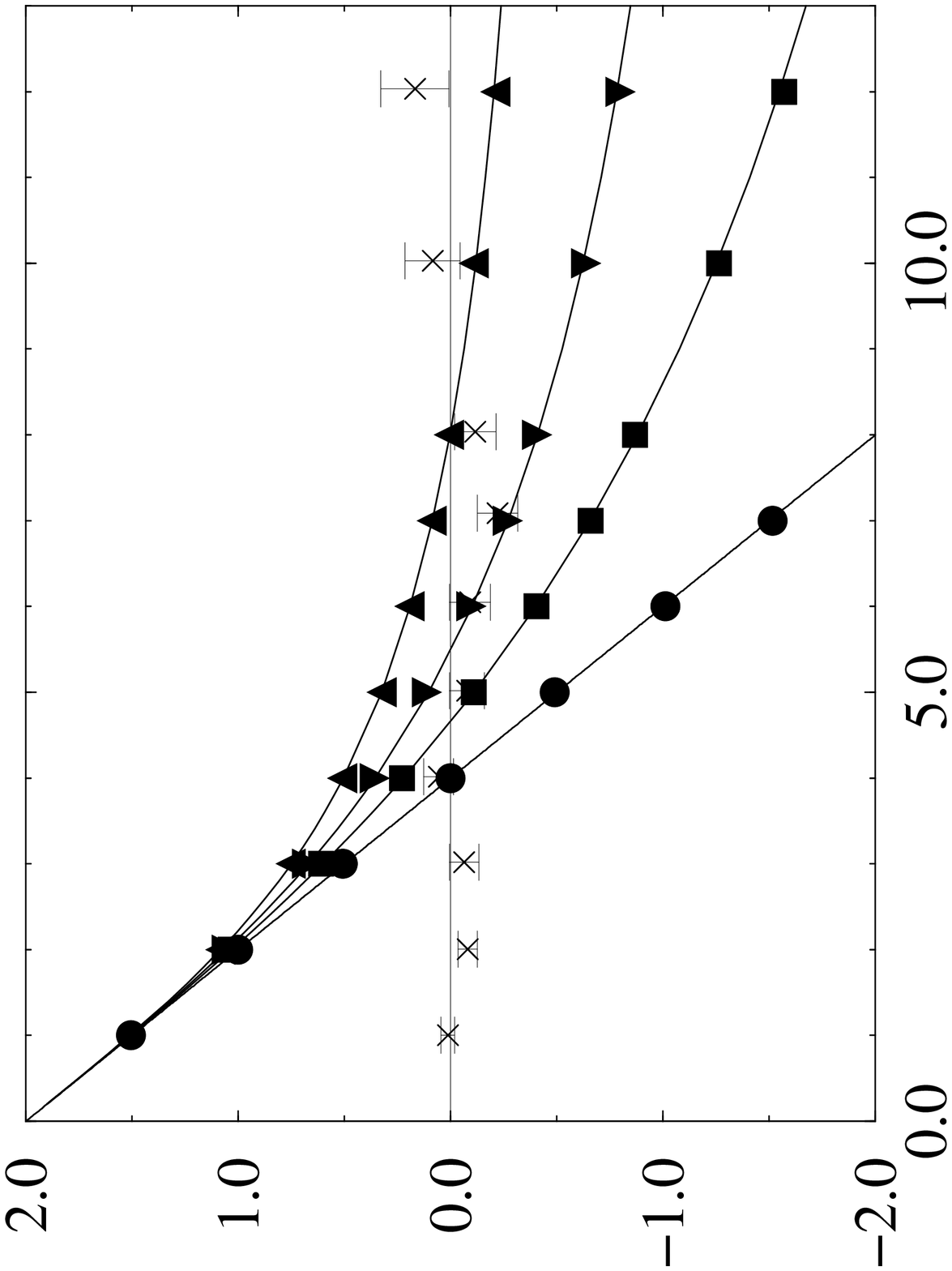}}}
 \put(1,125){\large $\delta/\lambda$}
 \put(65,54){\large$\lambda$}
\caption{The mean particle displacement $\delta/\lambda$ vs.\ $\lambda=\skl{l}$, cf.\ Eq.\ 
         (\ref{totaldelta}), shown for the model with $R=2$
         and different ES barriers.  Symbols represent
         the result of simulations with $L=1000$ for
         $p_{es}= 0 (\mbox{\Large$\bullet$}), 0.1 (\blacksquare), 0.2 
         (\mbox{\large $\blacktriangledown$}), 
         0.4
         (\mbox{\large $\blacktriangle$})$.
         Averages were performed over 100 runs and standard error bars
         would be smaller than the symbols, where not shown. 
         In addition, the correlation coefficient $\rho\!=\!(\skl{l_a l_b}\!-\!
         \skl{l}^2)/(\skl{l^2}\!-\!\skl{l}^2)$ for neighboring terraces $a$ and $b$
         is displayed for the case $p_{es}=0.2$;
         crosses and error bars correspond to $10\cdot\rho$.}
\label{strom}
\end{picture}
\end{center}
\end{figure}

 Since $\lc$ can only increase at the cost of shortening $\lr$ at the 
 same time, one obtains immediately the result
 \beq \label{lengthen}
   \tw (\lc\!\to\!\lc\!+\!1) = \left\{ \begin{array}{cl}
   \frac{1}{L} \klm{\lr\!-\!\Delta(\lr) \!+\!\Delta(\lc)} & \mbox{if~}
    \lr>0 \\
     0 & \mbox{if~} \lr=0. \end{array} \right.
   \eeq 
  We proceed by assuming that in a {\it population} of terraces the
 distribution of their sizes factorizes, i.e.\ that a single, 
identical
  $p(l)$ is sufficient to describe their statistics. 
 For the limiting case of an infinite ES barrier ($p_{es}=0$) the 
 evolution of a terrace is independent of the entire configuration 
 left of it, and the above property can be shown to hold true.
 In general, neighbor terraces clearly 
 interact. Nevertheless, our simulations show that the assumption of
 identically distributed, independent terrace sizes yields excellent
 approximations, at the very least. Figure \ref{strom} shows, for instance, that
  the correlation coefficient of neighboring terrace sizes in a 
  system with $L=1000$ and $p_{es} = 0.2$ vanishes within error bars.

 The analysis simplifies significantly if terrace sizes are considered
 to be uncorrelated.  The above expressions 
(\ref{shorten},\ref{lengthen}) were obtained for a given
triple of terraces $\gkl{\ell,c,r}$. By averaging
$\tw (\lc\!\to\!\lc\!\pm\!1)$ 
  over $p(\li)$ and
$p(\lr)$, respectively, one obtains the mean probabilities
$w(\lc\!\to\!\lc\!\pm\!1)$:
\begin{eqnarray}
 \label{lengthenmean}
  w(\lc\!\to\!\lc\!-\!1) & = & 
 %   \sum_{\lr =0}^{\infty} p(\lr) \tw(\lc\!\to\!\lc\!+\!1) = 
   \left\{ \begin{array}{cl} 
   \frac{1}{L} \klm{\lc\!-\!\Delta(\lc) \!+\!\skl{\Delta}} 
    & \mbox{for~~} \lc >0 \\
    0 & \mbox{for~~} \lc=0 \end{array} \right. \\
   \label{shortenmean}
   w (\lc\!\to\!\lc\!+\!1) & = &
     % \sum_{\li =0}^{\infty} \, p(\li) \tw (\lc \to \lc-1) \, = 
     \klm{\skl{l} \!-\!\skl{\Delta} \!+\! [1\!-\!p(0)]
    \Delta(\lc) } / L
\end{eqnarray}
where the r.h.s.\ involve only the width of the considered terrace itself,
the frequency $p(0)$ of vanishing terrace sizes and the mean values
 $\skl{l}=\sum_{j=0}^{\infty} \, j p(j)$ and $ \skl{\Delta} = 
 \sum_{j=0}^{\infty} \, \Delta(j) p(j)$, see Eq.\ (\ref{deltadef}).

 The evolution of terraces according to 
 (\ref{lengthenmean},\ref{shortenmean}) produces a stationary distribution $p(l)$, 
 if $ p(l\!+\!1) w(l\!+\!1\!\to\!l) = p(l)  w(l\!\to\!l\!+\!1) $,
 hence 
 \begin{equation} \label{recursion}
   p(l+1)  =  p(l) \, \frac{\skl{l} - \skl{\Delta} + [1\!-\!p(0)] \Delta(l)}
                           {l+1 - \Delta(l+1) + \skl{\Delta}} 
 \end{equation}
 for $l\geq0$. This relation is implicit, since all $p(l)$ have to be known
 for the evaluation of the averages on the r.h.s. 
 In the  particular case of an infinite ES barrier, Eq.\ (\ref{recursion})
 reads $ p(l\!+\!1)=p(l) \skl{l}/(l\!+\!1)$ which is  satisfied
 by the Poissonian $p(l) = \lambda^l \, e^{-\lambda} / l!$ with mean 
 $\skl{l}=\lambda$.   

 In order to obtain the stationary $p(l)$ for $p_{es}>0$ on a 
 surface with a given
 mean terrace size $\lambda$,  we  replace $\skl{l}$ with $\lambda$  
 and $\skl{\Delta}$ with an adjustable parameter $D$ in Eq.\ (\ref{recursion}). 
 The quantities $p(0)$ and $D$ are determined such 
 that $\sum_l p(l)=1$ is satisfied and $\skl{l}=\lambda$ is reproduced self--consistently.
 Note that then, by construction,
 (\ref{recursion})  guarantees $\skl{\Delta}=D$ as well. 
 In the numerical treatment, sums are truncated at a  value $l_{max}$, 
 with the resulting $p(l_{\max})$
 small enough to justify the truncation a posteriori.

 A particle deposited at, say, lattice site $i$ will become immobile at 
 a final position $j\neq i$ after incorporation and diffusion, in general. 
 The expected displacement $ \delta = \skl{j-i} $ depends on the distribution of
 terrace sizes in the system. Taking into account all possible displacement processes  
 and their corresponding probabilities one finds  
  \begin{eqnarray}  \label{totaldelta}
 \delta & = &
    \frac{1}{2} R (R+1) - \frac{1}{2} \sum_{l=0}^{R\!-\!1} p(l) \ekl{(R-l)(R-l+1)}  +  \\
  & & \sum_{l=R\!+\!1}^{\infty}  p(l) \klm{ -l(l\!-\!R\!-\!\Delta(l))\!+\!\frac{1}{2} l 
  (l\!+\!1) - \frac{1}{2} R(R\!+\!1)}, \nonumber 
  \end{eqnarray}
 where the $p(l)$ obtained from (\ref{recursion}) have to be inserted for a 
 given $\skl{l}=\lambda$. 
 Here, the first line corresponds to the expected (positive, downward)
 effect of incorporation 
 and the second represents the total (negative, upward) contribution of diffusion. 
 In the limiting case $p_{es}=0$ Eq. (\ref{totaldelta}) reduces to
 $\delta = R \skl{l} - \skl{l}^2/2$, exploiting the fact that $\Delta(l)=0$ in this case
 and $\skl{l^2} - \skl{l}^2 = \skl{l}$ for the Poisson distribution.

 Figure \ref{strom} shows the result of Monte Carlo simulations of 
 the growth process in a system of size $L=1000$ for the model with $R=2$ and
 different values of $p_{es}$, where  boundary conditions were used to fix  
 $\skl{l}=\lambda$.  We have displayed the particle current $\delta/\lambda$
 on the surface as a function of $\lambda$.
 Note that  $\delta$ in Eq. (\ref{totaldelta}) was obtained 
 for the normalization $\sum_l p(l)=1$
 which corresponds to a fixed number of terraces. 
 In systems with a fixed number $L$ of 
 lattice sites, an additional factor has to be introduced as the number of 
 terraces grows like $1/\lambda$.

 For very 
 steep surfaces, $\lambda\to0$, the displacement approaches
 the limiting value $R$,
 representing the fact that
\noindent
 every deposited particle is shifted by $R$ lattice
 constant in the incorporation and then comes to rest.  In the limit of vanishing
 slope, our model yields a diverging negative ``upward'' current. 
 This is an artifact of completely neglecting nucleation, which inevitably 
 becomes 
 important as $\lambda\to\infty$ and  imposes a maximal displacement on the order of $\ld$.

 On mounded surfaces, bottom and top terraces  limit the extension of
inclined flanks. Any slope that  results in a net uphill current according to 
 Eq.\ (\ref{totaldelta})  will steepen
 this portion of the surface and
vice versa. Accordingly, a mean terrace width $\lambda_o$ will be stabilized which corresponds
 to the zero of $\delta (\lambda)$. 
In the presence of an infinite ES--barrier we find the exact relation $\lambda_o=2R$.
A naive and not quite correct argument was used  in \cite{epl} to obtain
the same result.
It is instructive to check that the magic slope cannot be obtained
from the condition that the mean displacement vanishes on a particular terrace of size
$\hat{\lambda}$. This would correspond to setting $p(l)=\delta_{l,\hat{\lambda}}$ 
in Eq. (\ref{totaldelta})
and ~ gives results ~ analogous to the BCF--like 
treatment in \cite{georg} 
which does not account for fluctuating terrace sizes. 
For $p_{es}=0$ one obtains, e.g., $\hat{\lambda}=2R\!+\!1\!=\!\lambda_o\!+\!1$.     

\begin{figure}[t]
\begin{center}
\setlength{\unitlength}{2pt}
\begin{picture}(160,108)(0,0)
\put(0,0){\makebox(160,105)
          {\includegraphics{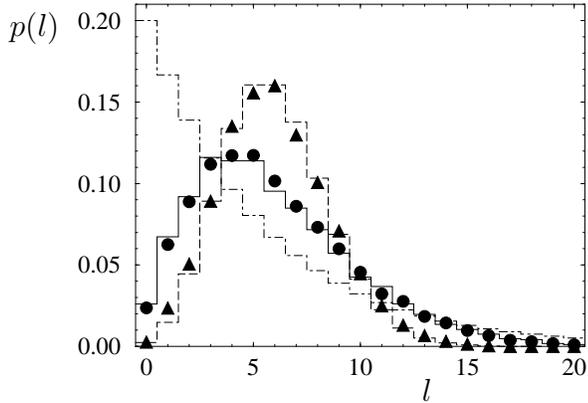}}}
 \put(2,105){\large $p(l)$}
 \put(80,36){\large$l$}
\caption{Frequency $p(l)$ of terrace sizes in two cases with 
$\skl{l}=\lambda_o=6$. The dashed line shows the theoretical
 prediction for $R=3, p_{es}=0$, triangles represent
simulations (as in Fig.\ \ref{strom}); the solid line and bullets
correspond to $R=2, p_{es}=0.258$.
The dot--dashed curve displays a geometric distribution with $\skl{l}=6$
for comparison.}
\label{distri}
\end{picture}
\end{center}
\end{figure}

\begin{figure}[t]
\begin{center}
\setlength{\unitlength}{2pt}
\begin{picture}(160,108)(0,0)
\put(0,0){\makebox(160,105)
          {\includegraphics{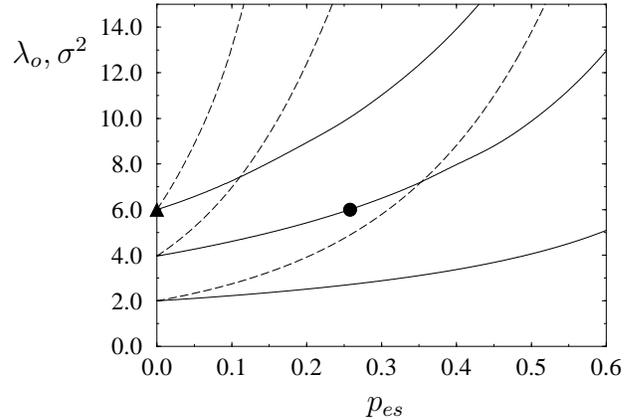}}}
 \put(1,100){\large $\lambda_o,\sigma^2$}
 \put(68,34){\large$p_{es}$}
\caption{The selected mean $\lambda_o$ (solid lines) and $\sigma^2 =
\skl{l^2}-\lambda_o^2$ (dashed) vs.\ $p_{es}$.
Pairs of curves correspond to (from below) $R=1,2,3$.
Note that for $p_{es}=0$ we find $\lambda_o=\sigma=2R$. The symbols represent  two choices
of $(p_{es},R)$ which result in $\lambda_o=6 $, cf.\ Fig.\ \ref{distri}.}
\label{mvsp}
\end{picture}
\end{center}
\end{figure}

Fig. \ref{distri} shows the frequency of terrace sizes as observed in two different
systems which both stabilize the mean $\lambda_o=6$. 
In one we have set $p_{es}=0$
and $R=3$, the second \mbox{example corresponds to $R=2$ and $p_{es}=0.258$.}
Computer simulations
show excellent agreement. 
Systems with a very pronounced ES--effect produce a narrow distribution
with a very low frequency $p(0)$ of {\sl step\/}
{\sl bunching\/}, i.e. zero terrace sizes.
As a limiting case one
finds $p(0)=e^{-\lambda_o}=e^{-2R}$ for infinite ES--barrier $(p_{es}=0)$. 
\noindent
On the
contrary, step bunching is observed with a much larger frequency in cases
with a weaker ES--effect where the distribution $p(l)$ is much broader.
Fig.\ \ref{mvsp} displays $\lambda_o$ and the variance $\sigma^2$ of the 
terrace size distribution as functions of $p_{es}$. Note that $\sigma^2$ 
grows drastically with increasing $p_{es}$, indicating 
significant deviations from the Poissonian for
infinite ES--effect.  

The analysis of experimental data is frequently based on the
simple assumption of random, non--interacting terrace sizes on 
vicinal surfaces. This leads to the geometric distribution 
$p_g(l) = (1-1/\lambda)^{l-1} / \lambda $ with $\skl{l}=\lambda$, 
see e.g.\ \cite{wollschlaeger}  for a discussion. 
Note that $p_g(l)$ differs significantly from the type of statistics 
that we find in our model, cf.\ Fig.\ \ref{distri}. 
In particular, step bunching is much more frequent in this
simple picture: $p_g(0)=(\lambda\!-\!1)^{-1}$.

In summary we have presented a  microscopic model of unstable 
epitaxial growth in which it is possible to derive the net 
particle currents on surfaces of a given inclination.  For the
first time it is possible to work out the full distribution of 
terrace sizes in such a system. Further,
we were able to calculate the stable mean terrace 
size and the corresponding statistical properties of the surface. 
We have restricted ourselves to the 
analysis of $(1+1)$--dimensional growth in this work. 
However, our results should carry over to a more 
realistic $(2+1)$--dimensional picture to a large extent, 
whenever terrace edges do not meander significantly.

Our findings allow for a qualitative interpretation of experimental results
in systems which display slope selection: frequent step bunching
and a broad distribution  hint at a relatively weak ES--barrier.  
Narrow distributions with little or no step bunching indicate that
a significant ES--effect is present but is compensated for 
by smoothening effects like downhill funneling. 

Extensions of this work  will concern  desorption and its
\noindent
 influence on the growth process. Preliminary results indicate that
a significant desorption rate can trigger a transition from slope selection
to rough growth
and we expect  non--trivial 
effects in the statistics of terrace sizes. 

This work was supported by the Deutsche Forschungsgemeinschaft
through a grant (M. Ahr) and through the 
Sonderforschungsberich 410 (S. Schinzer). We thank R. Metzler
 for a critical reading of the manuscript.

\ \\[-1.0cm]

\end{document}